\documentclass[letter]{aa} 
\usepackage{graphicx}
\usepackage{txfonts}
\begin{document}
\title{Herschel Observations of the W43 ``mini-starburst"
\thanks{{\it Herschel} in an ESA space observatory with science
instruments provided by European-led Principal Investigator
consortia and with important participation by NASA.}}
\author{J.  Bally\inst{1} \and
                 L. D. Anderson\inst{2} \and
                 C.  Battersby\inst{1} \and
                 L. Calzoletti \inst{3} \and
                 A. M.  DiGiorgio\inst{4} \and 
                 F. Faustini\inst{3} \and 
                 A.  Ginsburg\inst{1} \and 
                 J.  Z.  Li\inst{5} \and
                 Q. Nguyen-Luong\inst{6} \and
                 S.  Molinari\inst{4} \and
                 F.  Motte\inst{7} \and
                 M.  Pestalozzi\inst{8}  \and
                 R.  Plume\inst{9} \and
                 J.  Rodon\inst{2} \and
                 P.  Schilke\inst{10} \and 
                 W.  Schlingman\inst{11} \and 
                 N. Schneider-Bontemps\inst{12} \and
                 Y. Shirley\inst{11} \and
                 G.  S. Stringfellow\inst{13} \and
                 L. Testi\inst{14} \and
                 A. Traficante\inst{15} \and
                 M. Veneziani\inst{16} \and
                 A.  Zavagno\inst{2}        
          }
 \institute{Center for Astrophysics and Space Astronomy,  
   Department of Astrophysical and Planetary Sciences, 
      University of Colorado, UCB 389 
      Boulder CO 80309-0389, 
      USA
              \email{john.bally@colorado.edu}
 \and
      Laboratoire d'Astrophysique de Marseille (UMR 6110 CNRS \&  
      Universit\'e de Provence), 38 rue F. Joliot-Curie,  13388 Marseille Cedex 13, France 
  \and
      ASI Science Data Center, I-00044 Frascati (Rome), Italy 
  \and
      Istituto Fisica Spazio Interplanetario   
      INAF, via Fosso del Cavaliere 100, 00133 Roma, 
      Italy
  \and
      National Astronomical Observatories,    
      Chinese Academy of Sciences, Beijing 100012, 
      China
  \and
      Laboratoire AIM, CEA/IRFU Ð CNRS/INSU Ð Universite Paris  
      Diderot, CEA-Saclay, F-91191 Gif-sur-Yvette Cedex, France
 \and
      Laboratoire AIM, CEA/DSM - CNRS   
     UniversitŽ Paris Diderot, 
     DAPNIA/Service d'Astrophysique, B‰t. 709, 
     CEA-Saclay, 91191 Gif-sur-Yvette Cedex, 
     France
  \and
      Dept. of Physics,     
      University of Gothenburg, 
      412 96, G\"oteborg, 
      Sweden
  \and
      Department of Physics and Astronomy,    
      University of Calgary, 2500 University Drive NW, 
      Calgary, AB T2N 1N4, 
      Canada
 \and
      I. Physikalisches Instiut der Universit\"at zu K\"oln, Z\"ulpicher Str.   
      77, 50937 K\"oln, Germany
 \and
      Steward Observatory, University of Arizona,  933 North Cherry Ave.,   
      Tucson, AZ 85721
 \and
      SAp-CEA/Saclay, 91191 Gif-sur-Yvette Cedex, France    
 \and
      Center for Astrophysics and Space Astronomy,     
      University of Colorado, UCB 389 CASA, 
      Boulder CO 80309-0389, 
      USA
 \and
      European Southern Observatory,        
      Karl Schwarzschild str. 2, 85748 Garching, 
      Germany
  \and
      Dipartimento di Fisica, Universitˆ di Roma 2 "Tor Vergata", Rome,
      Italy 
  \and
       Dipartimento di Fisica, Universitˆ di Roma 1 "La Sapienza", Rome,
      Italy 
              }

   \date{Received 31 March 2010 }

 
  \abstract
   {}
   {To explore the infrared and radio properties of one of the closest
     Galactic starburst regions.}
   {Images obtained with the \textit{Herschel Space Observatory} at wavelengths
    of 70, 160, 250, 350, and 500 $\mu$m using the PACS and SPIRE 
    arrays are analyzed and compared with
    radio continuum VLA data and 8 $\mu$m images from the  {\it Spitzer  
    Space Telescope}.   The morphology of the far-infrared emission is combined 
    with radial velocity measurements of  millimeter and centimeter wavelength 
    transitions to identify features likely to be associated with the W43 complex. }
   {The W43 star-forming complex  is resolved into a dense cluster of
   protostars, infrared dark clouds, and ridges of warm dust heated by 
   massive stars.   The 4 brightest compact sources  with 
   L $> 1.5 \times 10^4$ L$_{\odot}$ embedded within the Z-shaped
   ridge of bright dust emission in W43 remain single  at 4\arcsec\  (0.1 pc) resolution 
   These objects,  likely to be massive protostars or compact clusters in early stages of 
   evolution are embedded in clumps with  masses of $10^3$ to $10^4$ M$_{\odot}$, but  
   contribute  only 2\% to the $3.6 \times 10^6$ L$_{\odot}$  far-IR luminosity of 
   W43 measured  in a 16 by 16 pc box.    The total mass of gas 
   derived from the far-IR dust emission inside this region is $\sim 10^6$ M$_{\odot}$.  
   Cometary dust clouds, compact 6 cm radio sources, and warm dust
   mark the locations of  older populations of massive stars.  Energy release has 
   created a cavity blowing-out below the Galactic plane.  Compression
   of  molecular  gas in the plane by the older H\,{\sc{ii}} region near G30.684--0.260  and
   the bipolar structure of the resulting younger W43 H\,{\sc{ii}} region may have triggered 
   the current mini-star burst.  }
   {}

   \keywords{Stars: protostars --
              Stars: massive --
              (ISM): HII regions --
              Infrared: ISM --
              ISM: W43
               }

   \maketitle
%

\section{Introduction}

The W43 "mini-starburst" region in the Molecular Ring near $l$=30.8$\degr$  
is one of the most luminous star forming complexes in the Galaxy
(Motte, Schilke, \& Lis 2003).  Located at a distance of
about 5.5 kpc at V$_{LSR} \approx$  85 to 107 km~s$^{-1}$, W43 contains a giant
H\,{\sc{ii}} region powered by a cluster of OB and Wolf-Rayet stars emitting a Lyman 
continuum luminosity of  about $10^{51}$ ionizing photons
per second (Smith, Biermann, \& Mezger 1978; Lester et al. 1985; Blum,
Damineli, \& Conti 1999).    The H\,{\sc{ii}} region  is in  contact with a 20 pc diameter
giant molecular cloud (GMC) with a mass  of about $10^6$~M$_{\odot}$ (Liszt 1995)and 
a total IR luminosity of  $\sim 3.5 \times 10^6$~L$_{\odot}$ (Lester 1985).   Motte
et al. (2003) identified about 50 clumps with masses ranging from 
40 to 4,000 M$_{\odot}$ in  350 and 1100 $\mu$m maps of the dust continuum.

W43 may be a "Rosetta Stone" for studies of  super-star cluster formation 
using Hi-GAL data. Comparison of the Hi-GAL data with ground based  
radio data and space based {\it Spitzer}  images show that on-going 
star formation is confined to a Z-shaped region abutting the W43 H\,{\sc{ii}} region.   
 

\section{Observations}

Images of four square degrees centered at $l$ = 30\degr , $b$ = 0\degr\ at wavelengths of 
70, 160, 250, 350, and 500 $\mu$m were obtained with the PACS (Griffin et al. 2010) 
and SPIRE (Poglitsch et al. 2010)  instruments 
on the 3.5 m diameter telescope  on  the {\it Herschel Space Observatory} (Pilbratt et al. 2008) 
as part of the Science  Demonstration Program  for the Hi-GAL Galactic Plane Survey  
(Molinari et al. 2010).   Each Hi-GAL image was  "unsharp-masked" to suppress diffuse 
Galactic emission by convolving  with a $\sigma$ = 70\arcsec\  gaussian to 
make a mask and subtracting this mask from the corresponding Hi-GAL image to 
enhance the visibility of  small-sacle structure ($<$ 200\arcsec\  $\sim$ 6 pc).

\section{Results}

\subsection{The W43 chimney:  A young superbubble?}

The {\it Herschel} images provide the first wide-field far-infrared views of the brightest
portion of the Galactic plane near  $l$ = 30\degr , allowing the investigation of the 
large-scale environment of the W43 mini-starburst complex (Figure 1).    W43 lies at the
top of the most prominent cavity in the four square-degree $l$ = 30\degr\  {\it Herschel} field.
The cavity consists of an S-shaped `hole'  in the dust emission bounded by a 6\arcmin\ radius 
curved ridge of dust extending from the left-side of W43 towards  G30.772$-$0.209 and 
continuing to  G30.684$-$0.260.   Towards low-longitudes (right side of Figure 1) the cavity is
bounded by a 0.25\degr -long vertical wall near $l$ = 30.5.  Most  of the cavity interior 
is filled with  faint free-free emission  in the  20 cm  MAGPIS VLA survey  and the 5 
GHz filled-aperture survey of  Altenhoff et al. (1979).    The cavity extends to at least 
Galactic latitude $b$ = --0.6 where it becomes confused with the foreground H\,{\sc{ii}} 
region Sh2-67  at  $V_{LSR}$ = 18 km~s$^{-1}$ (d $\sim$ 400 pc;
Fich, Treffers, \& Dahl 1990)  excited by BD-02 4752, a B0.5V star with visual 
magnitude 10.5.

The radial velocities of selected dust clumps were measured using emission
from  the  98 GHz CS 2$-$1 transition (Shirley et al. 2010) and with  $^{13}$CO 1$-$0 
data from the FCRAO Galactic Ring Survey  (Jackson et al. 2006).     In Figures 1 and 3,
CS radial velocities  with $85$ km~s$^{-1}  < V_{LSR} < 107$  km~s$^{-1}$
are shown in yellow; radial  velocities outside this range are shown in blue and cyan.    
Radial velocities based on $^{13}$CO 1$-$0 are shown in with smaller yellow circles.    
The clouds located below  W43, including 
the  6\arcmin\ radius ridge extending from the left side of W43 towards
G30.772-0.209  and the wall near $l$ = 30.5 are at  V$_{LSR}$ $\approx$ 
99 to 107 km~s$^{-1}$.   However, they appear to connect to the bright dust emission
in W43.   The larger radial velocities of these features compared to 
W43 may be due to acceleration by UV radiation and stellar winds.    Although 
line-of-sight confusion by unrelated  features is highly probable 
at $l$ = 30\degr ,  the preponderance of radial velocities similar to that of W43 make it 
likely that the Chimney is powered by massive young stars at a common distance of
about 5.5 kpc.   At this  distance,  the W43 Chimney is at least 70 pc long.    

Infrared dark clouds (IRDCs)  are seen in silhouette against background emission at
wavelengths below 70 $\mu$m but are bright at wavelengths  beyond 160 $\mu$m. 
The cometary IRDC  G30.772$-$0.209 and its bright rims  
(Figures 2 \&  3) point towards   G30.684$-$0.260 (IRAS 18456-0210). The 20 cm 
continuum   and  diffuse 24 $\mu$m  {\it Spitzer} emission (not shown)  indicate that 
massive stars  are located just above G30.684$-$0.260.    Figure 3 shows two 
concentrations of   6 cm mJy point-like radio sources in the MAGPIS survey 
(White et al. 2005;  Helfand et al. 2006) that trace free-free emission from 
compact H\,{\sc{ii}} regions,  ionized massive-star winds,  or the brightest compact 
features in  extended  H\,{\sc{ii}} regions  resolved by the VLA. The largest 
concentration is centered  on W43 and a second group is located a few arcminutes 
below  G30.684$-$0.260  around  the UC H\,{\sc{ii}}  region G30.667--0.332 
embedded in a $^{13}$CO cloud at V$_{LSR}$ = 89.4 km~s$^{-1}$.  While the
upper portion of the Chimney is clearly powered by W43, the southern portion appears
to contain an older group of massive stars which produce the compact 6 cm sources, 
a diffuse H\,{\sc{ii}}  region,  and several cometary clouds.

\subsection{The W43 starburst region}

The W43 region is shown in detail in Figures 2 and 3.
The 0.1\degr\ ($\sim$10 pc) long, Z-shaped W43 ridge of warm dust  (red lines 
in Figures  1 and 3) is the brightest source of far-IR, sub-mm, and mm emission in the 
$l$ = 30\degr\   field,  and one of the brightest in the entire Galaxy (Smith, Biermann,
\& Mezger 1978) .  At 70 $\mu$m, the Z-shaped ridge of bright dust emission contains a 
chain of about a  dozen compact sources  ($<$4\arcsec\ or 0.1 pc diameter) superimposed 
on the  bright background of warm dust associated  with the edges of the W43 H\,{\sc{ii}} 
region.

The Z-shaped ridge  is situated at the `waist' of a bipolar H\,{\sc{ii}} region 
that breaks-out towards both positive and negative Galactic latitudes. 
While the high surface-brightness lobe of W43 breaking out above the bar of the Z is 
confined to a 4 by 10 pc region,  below the Z the  lobe  extends to 
the cometary  clump G30.772$-$0.209  about  11\arcmin\ ($\sim$ 19 pc)  from the W43
central cluster.   
Filamentary 70 $\mu$m dust emission closely follows  the radio continuum emission at 
20 cm in the MAGPIS survey   and presumably
traces warm dust in photon-dominated regions (PDR) located just outside ionization fronts
illuminated by W43's massive O and W-R stars.    Ridges of colder dust seen absorption 
at 70 $\mu$m and below but in emission longward of 160 $\mu$m form arcs extending 
beyond the legs of the `Z' .   The curved ridge 
which peels off the high-longitude end of the `Z' is larger and  forms a semi-circle 
of dust with a  radius of about 6\arcmin\ (10 pc) and  may be responsible for confining 
the eastern  (low-latitude) lobe of the W43 H\,{\sc{ii}} region.   This arc 
terminates in a prominent cometary clump,  G30.772$-$0.209.   A second chain of 
IRDCs extends from MM3  and wraps around the 
high-latitude side of the H\,{\sc{ii}} region, blocking its expansion  in that direction.

\begin{figure}
\centering
\includegraphics[width=9cm]{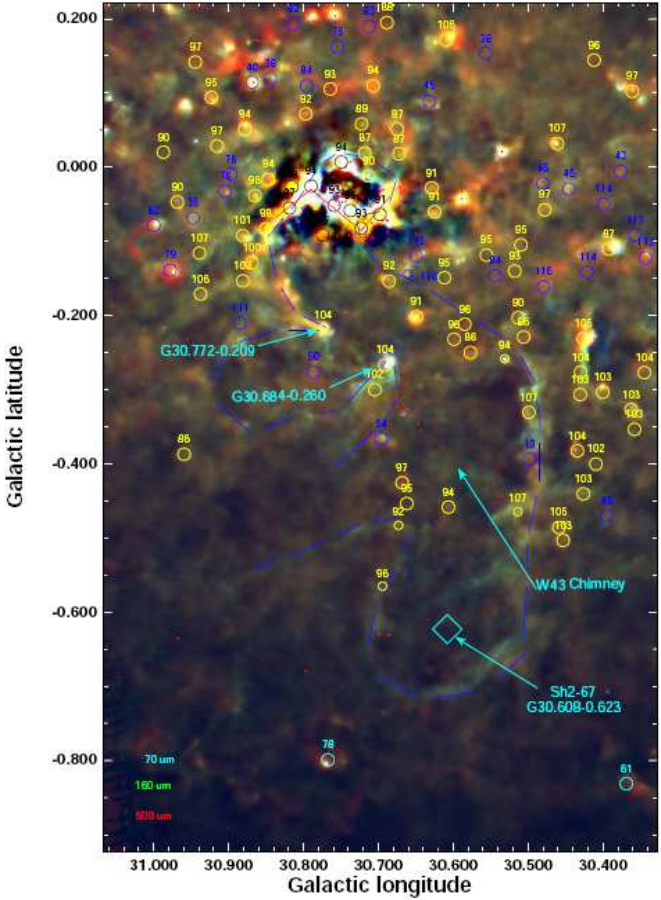}
\caption{
A color composite image showing the W43 starburst region
and the superbubble bursting towards low Galactic latitudes at 
70 $\mu$m (blue), 160 $\mu$m (green), and 500 $\mu$m (red) using
"unsharp-masked" images as described in the text .
Small red circles or dots mark the locations of 6 cm point sources.
Large yellow circles mark locations where the radial velocity has been measured 
using mm-wavelength tracers;  yellow  indicates radial velocities within the
velocity range 85 $< V_{LSR} < $ 107 km~s$^{-1}$ measured using 
CS  (Shirley et al. 2010, in prep.) or $^{13}$CO.  Blue and cyan circles indicate LSR velocities outside 
this range.  Small yellow circles mark the locations of $^{13}$CO clouds having similar 
radial velocities as W43.    The numbers above each circle mark the LSR radial 
velocity in km~s$^{-1}$.   Several regions  discussed in the text are
marked in cyan.  Warm dust at 70$\mu$m outlines the walls of the "W43 Chimney"
(shown in blue line segments).  A 0.1\degr\  interval corresponds to a linear
scale of 9.6 pc at the assumed 5.5 kpc distance to W43.
}
\label{fig1}%
\end{figure}

G30.720-0.083  (= MM3 in Motte et al. 2003; 115 mJy at 6 cm) is the brightest 
compact  H\,{\sc{ii}} region in the W43 complex.   It is centered  in a 1 pc radius 
IRDC  seen in  absorption at 70 $\mu$m and below  and as a  bright  clump 
beyond 160 $\mu$m.    Prominent  ionization fronts wrap around the side facing 
W43's central  cluster.    G30.817-0.056 (= MM1) is located at the head of 
parsec-long cometary IRDC that faces the central part of the W43 H\,{\sc{ii}} region 
seen in silhouette at wavelengths less than 70 $\mu$m.
MM1 is not detected at 6 and 20 cm  and may thus be in an earlier state of
evolution than MM3.  It  is likely a  massive proto-star or cluster of proto-stars
exhibiting OH, H$_2$O, and CH$_3$OH masers (see Motte et al. 2003).  
At  4\arcsec\  (0.1 pc)  resolution at 70 $\mu$m, this and the other luminous 
sub-mm sources remain single and unresolved,  implying that they are either 
isolated massive protostars, or clusters smaller than 0.1 pc.

The spectra of the four brightest compact sources, MM1 through 4 (Motte et al. 2003; 
marked in Figure 2) peak in the SPIRE 160 $\mu$m filter (Table 1).    These FIR 
sources contain  massive stars or compact clusters smaller than 0.1 pc in radius.  
The contribution of  mid-IR emission below 60 $\mu$m, and radiation that has 
escaped into the colder  extended envelope (such as the 1 pc radius envelope 
around MM3) may raise the total luminosity of each to nearly  $10^5$ L$_{\odot}$.

The unresolved emission 
from the 4 brightest MM sources  provide about 2\% of the total far-IR luminosity.   
The several dozen compact  sources (Motte et al. 2003 listed 51) in the Z-shaped ridge 
may contribute an additional   $\sim$ 6 \%.      These  compact sources probably represent 
massive stars with L $> 10^4$ L$_{\odot}$ in various stages of formation.  
MM1 is probably the youngest since it is the most luminous in the far-IR, is associated 
with masers,   but lacks  free-free emission.    MM3 is more evolved since it is associated  
with the brightest  compact H\,{\sc{ii}} region at 6 cm, but still highly embedded within 
the  Z-shaped ridge.     Masses for the brightest compact sources are
estimated using the 500 $\mu$m fluxes since the sources are most likely to be optically
thin and on the Rayleigh-Jeans tail of the spectrum.  We use the dust opacities from
Ossenkopf \& Henning (1994; OH94)  for the bracketing cases of MRN grains without and
with  thin ice mantles  evolved for $10^5$ years at densities of $10^6$ cm$^{-3}$.  Dust 
temperatures listed in Table~1 were determined using  grey body fits to the SEDs.   
The best-fit Robitaille (2006) models are consistent with the envelope masses in Table~1, 
imply stellar masses around 10 to 23 M$_{\odot}$ accreting at rates of around 
$10^{-3}$ M$_{\odot}$~yr$^{-1}$.  The beam-averaged envelope column densities 
range from N(H$_2$) $\sim 2 \times 10^{22}$ cm$^{-2}$ (MM4) to
$\sim 1.6  \times 10^{22}$ cm$^{-2}$ (MM1).

The Hi-GAL data provides the best estimate of the total luminosity of the W43 complex
which peaks around 160 $\mu$m, implying a typical dust temperature of around 20 K.
Summing the fluxes in a 600\arcsec\ by 600\arcsec\ (16 pc) box centered on W43 
yields a total far-IR luminosity of $L_{tot} = 3.6 \times 10^6$ L$_{\odot}$, within 10\% of
previous estimates (Lester et al. 1985).     This is a lower limit since some radiation may 
escape to larger distances than the measurement box.  Presumably, most of this luminosity 
is produced by the central WR+O cluster the massive stars traced by the 6 cm point sources 
and is reprocessed into the far-IR by the  surrounding dust.     Summing the total flux at
500 $\mu$m in this box, subtracting the background be averaging the surrounding 
annulus,  assuming a grain temperature of 20 K,  and a gas:dust ratio of 100 implies a
total mass of $0.87 - 1.3 \times 10^6$~M$_{\odot}$ ($3.7 \times 10^6$~M$_{\odot}$ for
un-evolved MRN grains).   

WR stars are post-main sequence  states of the most massive stars
(M $>$ 60 M$_{\odot}$) that have main-sequence lives of about 4 to 6 Myr.  
Thus,  the age of the WR+O cluster in W43 is probably in this range.  
If massive stars  have been forming at a constant rate  over a 5 Myr period,
the ratio of the luminosity of the 4 brightest objects, MM1 through 4, divided by
the total luminosity of the region implies that the duration of the embedded
massive protostar stage for such objects is 
$t_{proto} \sim L_{proto} / L_{tot} < 10^5$ years.

   \begin{table}
      \caption[]{Properties of the Brightest Compact W43 Sources.}
         \label{table1}
     $$ 
         \begin{array}{lcccccccc}
            \hline
            \noalign{\smallskip}
            Source      &   S_{70}   &  S_{160}   &S_{250}   &S_{350}   &S_{500}   &   T(K) & L / {[\mathrm{10^4 L_{\odot}}]}   &    M / {[\mathrm{10^3 M_{\odot}}]} \\
            \noalign{\smallskip}
            \hline
            \noalign{\smallskip}
            MM1          &  128  &  1943   & 1000  &  340   & 136   & 25 & 3.0  &   2.6 -  4.0 \\
            MM2          &  154  &  1400   &   700  &  246   & 164   & 23  & 1.5  &  3.5-  5.3  \\
            MM3          &  277  &  1255   & (400)  & 153   &   89   &  27 & 2.0   & 1.5 - 2.3 \\
            MM4          &  670  &  1148   &   205   & 131   &  53    &  28  & 2.0  & 0.9  - 1.3 \\
            \noalign{\smallskip}
            \hline
         \end{array}
     $$ 
\begin{list}{}{}
\item[$^{\mathrm{a}}$]  Temperatures are determined from single grey body fits assuming an
emissivity $\beta$ = 2, and excluding the 70 $\mu$m data points.
\item[$^{\mathrm{b}}$]  Fluxes are in Jy.  Luminosities only include flux from 
60$\mu$m to 600$\mu$m and are based on fluxes within a Hi-GAL resolution element.  
\item[$^{\mathrm{c}}$]  Masses estimated from $\lambda$ = 500 $\mu$m in a 37\arcsec\ beam, 
assuming  OH94 opacities at $t = 10^5$ years, $n = 10^6$ cm$^{-3}$ for
thin ice mantles  (first number) and un-evolved MRN dust with no mantles (second number).
\end{list}
   \end{table}

\section{Discussion}

There are at least two locations in the `W43 Chimney' containing OB stars or 
associations;  W43 and G30.684$-$0.260.    Fich et al. (1990) have 
shown that G30.608$-$0.623 is likely to be associated with the foreground  H\,{\sc{ii}}
region Sh2-67.    However, many molecular clouds associated with the dust rim surrounding
this region have velocities similar to W43.    Thus, these clouds are likely to be associated
with the W43 chimney.   Based on the dim 20 cm continuum, the large sizes of cometary clouds
facing G30.684$-$0.260,  and their location up to 20 pc away,  the massive stars in this region 
(marked in part by the lower cluster of compact 6 cm sources in Figs. 1 and 3) must  to be older
than W43.     Ionization fronts up to 40 pc away along the $l$ = 30.5 wall  appear 
to be illuminated from this direction (Figure 1).     Assuming a typical propagations speed of 5 
km~s$^{-1}$ for a typical D-type I-front moving through the ISM, the age of this cluster is 
likely to be  5 to 10 Myrs.   We hypothesize that as the expanding  H\,{\sc{ii}} region powered by
this group impacted denser gas toward the Galactic plane,  it triggered the formation and 
gravitational collapse of the GMC that evolved into W43, giving birth to its oldest massive stars  
(the O and WR cluster).   Subsequently UV radiation from the central O+WR cluster 
compressed the parent cloud towards both low- and high Galactic  longitudes, triggering 
the formation  of additional massive stars. 

The bipolar morphology of the W43 H\,{\sc{ii}} region indicates that its evolution
has been constrained by dense gas associated with the warm  70 $\mu$m to 500  $\mu$m 
dust emission located in the horizontal central bar of the Z-shaped ridge.   As the
H\,{\sc{ii}} region expanded above and below the ridge, its pressure would have 
compressed it, possibly triggereing the  current `mini-starburst'  traced by the dozens of 
cores in the Z-shaped ridge.

The `W43 Chimney'  may be similar to but much younger than the superbubble 
emerging from the Orion OB association (Bally 2008) or the one powered by the IC 1805  cluster in 
the W4  complex (Basu et al. 1999).   The Lyman continuum radiation and total luminosity 
of  the W43 complex  implies that it  contains the equivalent of about 50
O7 stars.  When these stars  explode, they may cause the Chimney to blow out
of the Galaxy to drive a `galactic fountain'.

   \begin{figure}
   \centering
     \includegraphics[width=9cm]{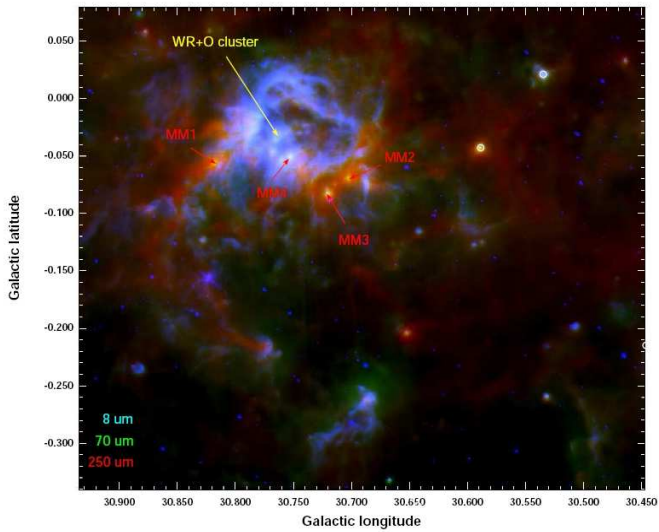}
        \caption{A color composite image showing the W43 starburst region
                      in  the {\it Spitzer} 8 $\mu$m (blue), and Hi-Gal 70 $\mu$m (green)
                      and 250 $\mu$m (red).  
                     }
        \label{fig2}%
      \end{figure}

   \begin{figure}
   \centering
     \includegraphics[width=9cm]{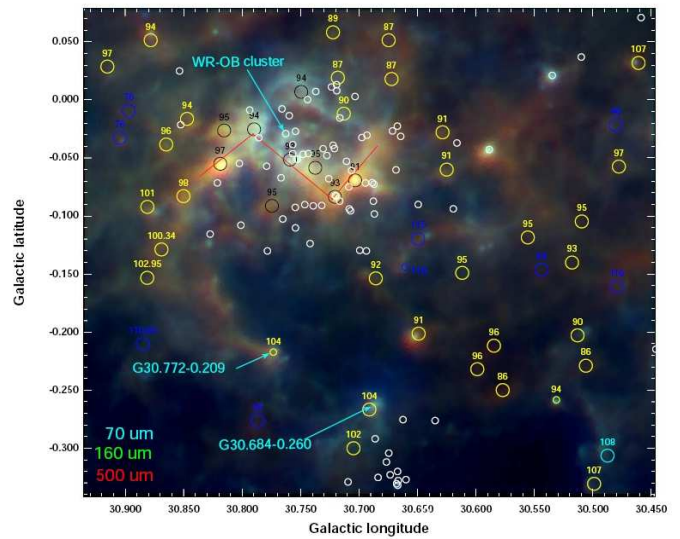}
        \caption{A color temperature image showing the W43 starburst region
                      at 70 $\mu$m (blue), 160 $\mu$m (green), 
                      and 500 $\mu$m (red).    Small  white 
                      circles mark the 6 cm point sources.  Black circles mark the brightest 
                      Motte et al. (2003) MM sources with their radial velocities.  As in Figure 1,
                      yellow, blue, cyan,  and  red circles mark locations where radial 
                      velocites have been determined; the number
                      above each circle give the LSR radial velocity.}
        \label{fig3}%
      \end{figure}

\section{Conclusions}

Hi-GAL provides the first high resolution far-infrared view of the Galactic ISM.    
W43 contains about a dozen embedded massive protostars with $L > 10^4$ L$_{\odot}$
which  contribute about 5\% to 8\% of the total luminosity, $L_{tot} = 3.6 \times 10^6$ L$_{\odot}$ 
measured in a 16 pc diameter box.   The youngest  luminous ($> 2 \times 10^4$  L$_{\odot}$) 
object, MM1 is  located at the head of a comet-shaped IRDC pointing away from the W43 central 
cluster.   MM3  is older since it is associated with the  brightest compact H\,{\sc{ii}} region in W43. 
Compression of the  Z-shaped filament by the bipolar W43 H\,{\sc{ii}} region
may have triggered the formation of dozens of luminous sources. 

The `W43 Chimney' is the most obvious giant cavity in the  Hi-GAL  fields and may 
represent a young 30 by 70 pc superbubble  filled with diffuse 20 cm emission and 
rimmed by  warm dust and molecular  clouds having similar  radial velocities as W43. 
Parsec-scale cometary clouds associated with G30.772-0.209 and G30.684-0.260 point 
to an older group of  massive stars below W43 and associated with a cluster of compact 
6 cm sources.  Compression of clouds  closer to the Galactic plane by this group
may have triggered the initial burst of star formation in W43.  W43 and this older group 
may be energizing the W43 Chimney.

\begin{acknowledgements}
The participation of J.B and G.S.S are supported in part by NASA through an 
award issued by JPL/Caltech via NASA Grant \#1350780.  We thank the 
referee for making excellent suggestions for improving the text.
\end{acknowledgements}


\begin{thebibliography}{}

\bibitem[Altenhoff et  al.(1979)]{1979A&AS...35...23A} 
    Altenhoff, W.~J., Downes, D., Pauls, T., \& Schraml, J.\ 1979, \aaps, 35, 23
    
\bibitem[Bally(2008)]{2008hsf1.book..459B} 
    Bally, J.\ 2008, Handbook of  Star Forming Regions, Volume I, 459

\bibitem[Balser et al.(2001)]{2001AJ....121..371B} 
     Balser, D.~S., Goss,  W.~M., \& De Pree, C.~G.\ 2001, \aj, 121, 371
\bibitem[]{}  
     Basu, S., Johnstone, D., \& Martin, P. G. 1999, ApJ, 516, 843

\bibitem[Blum et al.(1999)]{1999AJ....117.1392B} 
     Blum, R.~D., Damineli, A.,   \& Conti, P.~S.\ 1999, \aj, 117, 1392
     
\bibitem[]{}
    Fich, M., Treffers, R. R., \& Dahl, G. P.  1990, AJ, 99, 622
    
\bibitem[]{} 
   Griffin,  M. et al. 2010, A\&A, (in press).
   
\bibitem[]{} 
    Helfand, D.J., Becker, R.H., White, R.L., Fallon, A., \&  Tuttle, S. 2006,
    AJ, 131, 2525
     
\bibitem[]{}
     Jackson, J.~M. et al. 2006, ApJS, 163, 145
     

\bibitem[Lester et al.(1985)]{1985ApJ...296..565L}
      Lester, D.~F., 
      Dinerstein, H.~L., Werner, M.~W., Harvey, P.~M., Evans, N.~J., II, 
      \& Brown, R.~L.\ 1985, \apj, 296, 565
      
\bibitem[]{}
   Liszt, H. S. 1995, AJ, 109, 1205
      
\bibitem[]{}
  Molinari, S., \& the Hi-GAL Team, 2010, arXiv,arXiv:1000.2106

\bibitem[Motte et al.(2003)]{2003ApJ...582..277M} 
    Motte, F., Schilke, P.,  \& Lis, D.~C.\ 2003, \apj, 582, 277
      
\bibitem[]{} 
    Ossenkopf, V. \& Henning, T. 1994, A\&A, 291, 943
      
\bibitem[]{} 
   Pilbratt, G. L. et al. 2008, SPIE, 7010, 1
   
\bibitem[]{} 
   Poglitsch, A. et al. 2010, A\&A, (in press).
   
\bibitem[]{} 
    Robitaille, T.~P.,   Whitney, B.~A., Indebetouw, R., Wood, K., 
      \& Denzmore, P.\ 2006, \apjs, 167, 256 
        
\bibitem[Smith et al.(1978)]{1978A&A....66...65S} 
      Smith, L.~F., Biermann, P., \& Mezger, P.~G.\ 1978, \aap, 66, 65
      
\bibitem[]{}   
 White, R.L., Becker, R.H., and Helfand, D.J. 2005, AJ 130, 586
  
 \end{thebibliography}
\end{document}